\documentclass[compsoc,conference,a4paper,10pt,times]{IEEEtran}
\IEEEoverridecommandlockouts
\usepackage{cite}
\usepackage{amsmath,amssymb,amsfonts}
\usepackage{algorithmic}
\usepackage{graphicx}
\usepackage{textcomp}
\usepackage{bmpsize}
\usepackage{xcolor}
\usepackage[colorlinks=true,urlcolor=black]{hyperref}
\def\BibTeX{{\rm B\kern-.05em{\sc i\kern-.025em b}\kern-.08em
    T\kern-.1667em\lower.7ex\hbox{E}\kern-.125emX}}

\usepackage[binary-units=true]{siunitx}
\DeclareSIUnit[per-mode=symbol]\bps{\bit \per \second}
\usepackage{tikz}
\usetikzlibrary{calc,arrows,patterns}
\usepackage{colortbl}
\usepackage{bytefield}
\usepackage{pgfplots}
\pgfplotsset{compat=1.12}
\usepackage{pgfplotstable}
\usepackage{multirow}
\usepackage{booktabs}
\usetikzlibrary{decorations.pathreplacing}
\usepackage{todonotes}
\usepackage{csquotes}
\usepackage[caption=false, font=footnotesize]{subfig}
\definecolor{kitgreenexcl}{cmyk}{1.0,  0.0,  0.6, 0.0}
\definecolor{kitblue}     {cmyk}{0.8,  0.5,  0.0, 0.0}
\definecolor{kitgreen}    {cmyk}{0.6,  0.0,  1.0, 0.0}
\definecolor{kityellow}   {cmyk}{0.0,  0.05, 1.0, 0.0}
\definecolor{kitorange}   {cmyk}{0.0,  0.45, 1.0, 0.0}
\definecolor{kitbrown}    {cmyk}{0.35, 0.5,  1.0, 0.0}
\definecolor{kitred}      {cmyk}{0.25, 1.0,  1.0, 0.0}
\definecolor{kitpurple}   {cmyk}{0.25, 1.0,  0.0, 0.0}
\definecolor{kitcyan}     {cmyk}{0.9,  0.05, 0.0, 0.0}
\newcommand{\code}[1]{{\tt \small{#1}}}
\usepackage{listings}
\makeatletter
\newcommand{\linebreakand}{%
\end{@IEEEauthorhalign}
\hfill\mbox{}\par
\mbox{}\hfill\begin{@IEEEauthorhalign}
}
\makeatother

\DeclareCaptionType{listing}[Listing][List of Code Listings] 

\begin{document}

\title{TurboCC: A Practical Frequency-Based Covert Channel With Intel Turbo Boost}

\author{\IEEEauthorblockN{Manuel Kalmbach}
\IEEEauthorblockA{\textit{Karlsruhe Institute of Technology} \\
Karlsruhe, Germany \\
manuel.kalmbach@kit.edu}
\and
\IEEEauthorblockN{Mathias Gottschlag}
\IEEEauthorblockA{\textit{Karlsruhe Institute of Technology} \\
Karlsruhe, Germany \\
os@itec.kit.edu}
\and
\IEEEauthorblockN{Tim Schmidt}
\IEEEauthorblockA{\textit{Karlsruhe Institute of Technology} \\
	Karlsruhe, Germany \\
	os@itec.kit.edu}
\linebreakand 
\IEEEauthorblockN{Frank Bellosa}
\IEEEauthorblockA{\textit{Karlsruhe Institute of Technology} \\
	Karlsruhe, Germany \\
	os@itec.kit.edu}
}

\maketitle

\begin{abstract}
	Covert channels are communication channels used by attackers to transmit information from a compromised system when the access control policy of the system does not allow doing so.
	Previous work has shown that CPU frequency scaling can be used as a covert channel to transmit information between otherwise isolated processes.
	Modern systems either try to save power or try to operate near their power limits in order to maximize performance, so they implement mechanisms to vary the frequency based on load.
	Existing covert channels based on this approach are either easily thwarted by software countermeasures or only work on completely idle systems.

	In this paper, we show how the automatic frequency scaling provided by Intel Turbo Boost can be used to construct a covert channel that is both hard to prevent without significant performance impact and can tolerate significant background system load.
	As Intel Turbo Boost selects the maximum CPU frequency based on the number of active cores, our covert channel modulates information onto the maximum CPU frequency by placing load on multiple additional CPU cores.
	Our prototype of the covert channel achieves a throughput of up to \SI{61}{\bps} on an idle system and up to \SI{43}{\bps} on a system with 25\% utilization.
\end{abstract}


\section{Introduction}

Businesses often depend on secure storage of sensitive information.
To prevent inadvertent or malicious transfer of this information into the public, access control mechanisms such as firewalls or separate namespaces in a virtualized environment are used to restrict communication from sensitive software components.
In such a system, even if an attacker has gained control over sensitive software components, the attacker is not able to use the regular communication channels such as network or inter-process communication to access the information.

Previous work, however, has identified a wide range of \emph{covert channels} which are not covered by traditional access control.
A covert channel transmits information via a method not intended for information transfer~\cite{lampson_note_1973}.
Covert channels frequently employ hardware implementation artifacts such as memory access timing~\cite{maurice_hello_2017,wu2014whispers,pessl_drama_2016}, CPU temperature~\cite{masti_thermal_nodate} or power management characteristics~\cite{khatamifard_powert_2019} for communication. 
In the process, increasingly stronger isolation mechanisms such as cache partitioning~\cite{wang_covert_2006} or memory bandwidth partitioning~\cite{gundu2014memory} have been developed to prevent unauthorized flow of information.

Recently, the CPU frequency has been identified as another mechanism to construct covert channels~\cite{alagappan_dfs_2017,miedl_frequency_2018,khatamifard_powert_2019}.
In order to reduce power consumption, operating systems often reduce the CPU frequency as soon as the CPU becomes partially or fully idle.
Therefore, a process can modulate the load of the system to trigger intentional frequency changes.
If another process is able to read the CPU frequency, the two processes can use this mechanism to exchange information even if they are otherwise not permitted to do so.

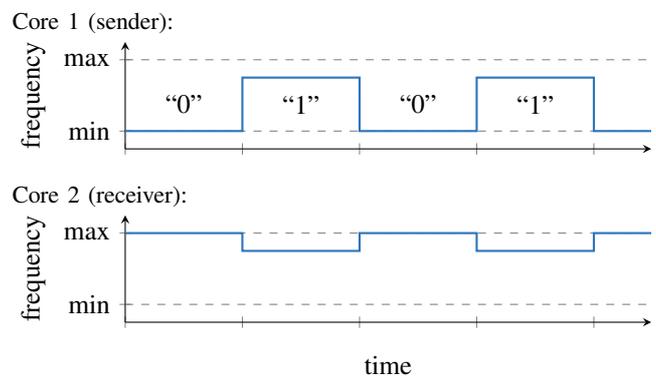
\begin{figure}[t]
	{\small Core 1 (sender):}\\
	\begin{tikzpicture}
	\begin{axis}[
	ylabel=frequency,
	axis x line = bottom,axis y line = left,
	xticklabels={,,},
	ytick={0.2,1.0},
	yticklabels={min,max},
	ymin=0,
	ymax=1.2,
	width=8.5cm,
	height=3cm,
	]
	
	\coordinate (minleft) at (axis cs:0,0.2);
	\coordinate (minright) at (axis cs:4.49,0.2);
	\draw [gray,sharp plot,dashed] (minleft) -- (minright);
	
	\coordinate (maxleft) at (axis cs:0,1);
	\coordinate (maxright) at (axis cs:4.49,1);
	\draw [gray,sharp plot,dashed] (maxleft) -- (maxright);
	
	\node[] at (0.5,0.5) {\enquote{0}};
	\node[] at (1.5,0.5) {\enquote{1}};
	\node[] at (2.5,0.5) {\enquote{0}};
	\node[] at (3.5,0.5) {\enquote{1}};
	
	\addplot+[const plot, no marks, thick, color=kitblue] coordinates {(0,0.2) (1,0.8) (2,0.2) (3,0.8) (4,0.2) (4.49,0.2)};
	\end{axis}
	\end{tikzpicture}
	{\small Core 2 (receiver):}\\
	\begin{tikzpicture}
	\begin{axis}[
	xlabel=time,
	ylabel=frequency,
	axis x line = bottom,axis y line = left,
	xticklabels={,,},
	ytick={0.2,1.0},
	yticklabels={min,max},
	ymin=0,
	ymax=1.2,
	width=8.5cm,
	height=3cm,
	]
	
	\coordinate (minleft) at (axis cs:0,0.2);
	\coordinate (minright) at (axis cs:4.49,0.2);
	\draw [gray,sharp plot,dashed] (minleft) -- (minright);
	
	\coordinate (maxleft) at (axis cs:0,1);
	\coordinate (maxright) at (axis cs:4.49,1);
	\draw [gray,sharp plot,dashed] (maxleft) -- (maxright);
	
	\addplot+[const plot, no marks, thick, color=kitblue] coordinates {(0,1) (1,0.8) (2,1) (3,0.8) (4,1) (4.49,1)};
	\end{axis}
	\end{tikzpicture}
	
\caption{
	The maximum boost frequency depends on the number of active cores.
	Increased load on one core is visible as a frequency drop on other cores, which can be used to transmit information.
}
\label{fig:turboprinciple}
\end{figure}

Reading the CPU frequency of the current core of a process is simple, as the process can use timers to determine the execution time of a known number of operations~\cite{miedl_frequency_2018}.
Modern CPUs, however, decouple the frequency of individual cores~\cite{hackenberg2015energy}.
Therefore, the OS can prevent the covert channel by placing sensitive processes on isolated cores:
In that case, each process is only able to read the frequency of the core its currently running on, whereas sensitive processes only affect their own core's frequency.
In this situation, system call interfaces such as the sysfs file system in Linux can be used to read the CPU frequency of all cores~\cite{alagappan_dfs_2017}.
Again, the OS can trivially restrict access to such interfaces, thereby greatly limiting the viability of existing frequency-based covert channels in practice.

A property of these modern systems that introduces new cross-core covert channels is that they operate close to their thermal limits, so utilizing available power headroom is necessary to maximize performance.
This situation requires power management mechanisms to prevent exceeding power limits, for example, by decreasing the frequency of one core if another increases its power utilization.
Khatamifard et al.~\cite{khatamifard_powert_2019} show that this principle can be used to construct covert channels -- which they named POWERT channels -- and they calculate the potential throughput.
They demonstrate a covert channel on an Intel system, albeit without explaining the underlying frequency scaling implementation which causes the frequency changes.
In fact, their demonstration likely works because Intel Turbo Boost 2 varies the frequency based on the number of active cores as shown in Table~\ref{tab:turbo_frequenzen}.
As shown in Figure~\ref{fig:turboprinciple}, if the transmitter activates an additional core, the receiver is able to detect the resulting frequency drop.
As Khatamifard et al. only vary the power consumption of a single CPU core -- a \enquote{1} is sent by three active cores, a \enquote{0} by two active cores -- their prototype fails if just a single additional core is used by other processes\footnote{If the cooling solution is not able to sustain the specified turbo frequency, thermal throttling might cause frequency changes in this situation -- our work targets server systems, where that situation is unlikely.}.
Our experiments show that even the little additional CPU utilization caused by virtualization can cause significant throughput reduction in such a system.

In this paper, we describe the potential of adapting such covert channels to specific power management mechanisms.
We describe a covert channel specially for Intel Turbo Boost 2 and show how adaptations make the covert channel viable even with significantly higher background load as required for the use in real-world scenarios.
Our experiments show that the covert channel is usable at up to 37.5\% background CPU load which is close to the CPU utilization achieved by modern data centers~\cite{garraghan_analysis_2013}.

In our prototype, the transmitter activates or deactivates multiple cores synchronously to reduce or increase the frequency.
The receiver executes a timing loop on another core to determine the current turbo frequency and employs edge detection on the resulting signal to recover the transmitted information.
As this setup cannot prevent some of the transmission errors due to interference from other processes, our prototype implements a packet-based communication protocol with automatic retransmission of corrupt packets to ensure a robust communication channel.
On a system with an 8-core Intel server processor, our prototype provides a throughput of \SI{61}{\bps} on an idle system and of \SI{43}{\bps} on a system running at 25\% utilization.
Our covert channel is still able to transmit \SI{12}{\bps} at a background load of 37.5\%. 
These results show that CPU load alone is not a viable countermeasure against frequency-based covert channels.

The following sections are structured as follows:
As our covert channel depends on Intel Turbo Boost, Section~\ref{sec:turbo_boost_analysis} contains a description of the behavior of Turbo Boost.
Then, Sections~\ref{sec:covertchannel} and \ref{sec:implementation} describe the design principles and the implementation of our covert channel.
The following sections then contain an evaluation of our prototype at different amounts of system utilization (Section~\ref{sec:evaluation}) and a discussion of countermeasures, limitations, and the potential use of AMD Precision Boost to construct covert channels (Sections~\ref{sec:counter}, \ref{sec:discussion} and \ref{sec:amd_precision_boost}).
Finally, we present related work in Section~\ref{sec:relwork} and conclude in Section~\ref{sec:conclusion}.

\subsection{Threat Model}

As a threat model, we assume a scenario similar to other covert channel publications described in Section \ref{sec:relwork}. 
Hereby an attacker is able to execute arbitrary code in user-space, but access control prevents information flow. 
We imagine a scenario where one process has access to sensitive information, but access control prevents any further communication with other processes or via the network. 
Another process, instead, has internet access but no access to the sensitive information. 
We assume that the attacker was able to infiltrate both processes and execute arbitrary code in their context (in our case, receiver and transmitter code of the covert channel). 
We further assume that the attacker was not able to gain elevated privileges to circumvent the access control policies.
For our covert channel to work, the two processes have to run on the same physical CPU package -- albeit on arbitrary CPU cores -- and the overall background load has to be low enough as described in Section \ref{sec:eval:backgroundload}.

\section{Intel Turbo Boost}
\label{sec:turbo_boost_analysis}

Traditionally, systems employ two types of frequency scaling:
At low load, the operating system reduces the frequency to conserve energy, and at high load the CPU opportunistically increases the frequency above its nominal maximum frequency to improve performance whenever further thermal headroom and power supply capacity is available.
Similar to POWERT~\cite{khatamifard_powert_2019}, this work exploits the latter.

\begin{table}[!t]
	\centering
	\caption{Turbo frequencies of the Intel Xeon Silver 4108 processor for different numbers of active cores~\cite{wikichip_xeon_2019}}
	\label{tab:turbo_frequenzen} 
	
	\definecolor{green1}{RGB}{0,255,0}
	\definecolor{green2}{RGB}{85,255,0}
	\definecolor{green3}{RGB}{30,255,0}
	\definecolor{green4}{RGB}{200,255,0}
	
	\definecolor{yellow1}{RGB}{255,255,0}
	\definecolor{orange1}{RGB}{255,170,0}
	
	\definecolor{red1}{RGB}{255,85,0}
	\definecolor{red2}{RGB}{255,0,0}
	
	\begin{tabular}{c|c|c|c}
		\toprule
		\multirow{2}{*}{\shortstack{Base\\Frequency}} & \multicolumn{3}{c}{\shortstack{Turbo Frequency / Active Cores}}\\
		& 1, 2 & 3, 4 & 5, 6, 7, 8 \\
		\midrule
		\SI{1.8}{\giga\hertz} & \cellcolor{green1!30} \SI{3.0}{\giga\hertz} &  \cellcolor{green2!30} \SI{2.7}{\giga\hertz} &  \cellcolor{yellow1!30} \SI{2.1}{\giga\hertz} \\
		\bottomrule
	\end{tabular} 
\end{table}

Our prototype targets Intel Turbo Boost 2.0 which is found in most recent Intel CPUs that tries to increase the active cores' frequency as much as possible without violating either power or temperature limits~\cite{charles_evaluation_2009}.
Our work exploits the fact that the power consumption depends on the number of cores which are active, which is why Turbo Boost selects higher turbo frequencies when more cores are in a power-saving sleep state (C-state\,$\geq$\,C3)~\cite{intel2008turboboostwhitepaper}. 
Table~\ref{tab:turbo_frequenzen} lists the frequencies for the Xeon Silver 4108 processor, showing that depending on the active core count the frequency is automatically increased to up to \SI{3}{\giga\hertz} from a base frequency of \SI{1.8}{\giga\hertz}.

Covert channels using the frequency selection to transmit information require frequent frequency changes to provide high throughput.
In the Haswell microarchitecture, a central power control unit (PCU)~\cite{bhandaru2016providing} determines the highest possible turbo frequency once per millisecond, calculating the maximum CPU frequency from the temperature of the chip as well as the time spent at the different power limits~\cite{servermeile_intel_nodate}.
Although our prototype has been developed on an Intel Skylake-SP CPU, our measurements confirm the rate of these turbo frequency changes.

This high frequency switching rate is only viable because Turbo Boost is integrated into the CPU and is not controlled by the operating system.
The operating system only has the possibility to activate or deactivate Turbo Boost by setting the P-states~\cite{wamhoff_turbo_2014}. 
If P-state P0 is set, the boost frequencies are activated.
At any higher P-state Turbo Boost is deactivated and the cores' frequency is either defined by the given P-state or by Intel's speed shift technology~\cite{doweck_inside_2017}.  

\section{Frequency Covert Channel}
\label{sec:covertchannel}
With Intel Turbo Boost the activity of one core affects the maximum frequency on other cores, which enables a covert channel between different cores.
Exchanging information only requires the transmitter to change the maximum core frequency and the receiver to detect the changes.
As the maximum frequency is affected by a number of parameters such as the number of active cores, power limitations, temperature constraints, or even the type of instructions executed, the transmitter could be built using different mechanisms.
In our work, we focus on the number of active cores as it provides a simple mechanism to influence the frequencies on a well-cooled system, where reaching power or temperature limits might be difficult.

Activating a single core is not always enough, as shown in Table \ref{tab:turbo_frequenzen}.
We can see that Turbo Boost uses the same turbo frequency for multiple numbers of active cores. 
For example, at most two cores are allowed to be active to get the highest turbo frequency of \SI{3.0}{\giga\hertz}.
Similarly, with three or four active cores the turbo frequency drops to \SI{2.7}{\giga\hertz}.
In this range, if we assume one core to be used by a receiver process as described below, activating a single additional core does not always cause a frequency change, whereas two freshly activated cores always will.

To read the current turbo frequency, a receiver has to utilize its core enough to use the turbo frequency and is then able to measure its current frequency. 
Because the turbo frequency is identical for all cores, the receiver can determine the current turbo frequency from any core it is running on. 

Note that once more than four cores are already active, on our system no amount of additional active cores will cause further frequency reductions as the turbo frequency already is at the lowest \enquote{all-core turbo} level.
Therefore, the covert channel is limited to at least partially idle systems.
Some system utilization is acceptable, though, as long as less than four cores are active.

In all such cases the transmitter needs to ensure that its frequency changes are distinguishable from the background noise.
In particular, if the system utilization is not constant, the transmitter needs to trigger a lower frequency than can be caused by the other processes.
If a larger number of cores are synchronously activated by the transmitter, the resulting frequency swing is larger and the resulting frequency is guaranteed to be lower.
Therefore, a larger number of transmitter cores increases the amount of system utilization that can be tolerated by the covert channel.

\section{Implementation}
\label{sec:implementation}

To determine the potential of a covert channel specifically adapted to Intel Turbo Boost, we implemented a prototype called TurboCC to transmit information between processes and virtual machines on a Linux system.
Our implementation consists of two basic parts: The transmitter (described in Section~\ref{sec:impl:transmitter}) changes the turbo frequency by activating additional cores, and the receiver (described in Section~\ref{sec:impl:receiver}) measures the frequency and recovers the information.

Intermittently, other processes on the system will activate additional cores and can potentially cause transmission errors.
Therefore, the covert channel uses a communication protocol with automatic retransmission of corrupt packets as described in Section~\ref{sec:imp:communication_protocol}.
Unlike other approaches, we do not use error-correcting codes.
Instead, we only use a CRC checksum to detect corruption, a decision which is detailed in Section~\ref{sec:impl:error_correction}.

\subsection{Transmitter}
\label{sec:impl:transmitter}

As shown in Figure~\ref{fig:turboprinciple}, our transmitter uses a simple approach.
It transmits a \enquote{1}-bit by keeping one or more cores active and a \enquote{0}-bit by letting them sleep for a given time.
In case of a \enquote{1}-bit, the additional active cores will cause Turbo Boost to select a lower turbo frequency which can be detected by the receiver.
A similar principle, albeit with only one core which is activated, is implemented in the POWERT prototype~\cite{khatamifard_powert_2019}.

The number of cores activated by the transmitter needs to be chosen so that the cores deterministically trigger a specific frequency change that can be recognized by the receiver.
Table~\ref{tab:turbo_frequenzen} shows the frequency levels for a 8-core server system.
As already described in Section~\ref{sec:covertchannel}, the receiver keeps one core activated, then two additional cores are required to switch to the next lower frequency level assuming that the system is otherwise idle.
More cores are required if other processes generate additional constant or changing background load, as then another frequency level might have to be targeted.
Particularly, if more cores are used to send a \enquote{1}-bit than are at most held active by other processes, the resulting frequency change is distinguishable from background noise.
Note that sometimes fewer cores are sufficient if the transmitter monitors the resulting frequency and adds more cores only when required.
Doing so, however, potentially delays the frequency change and therefore reduces the throughput.
In our prototype, we make the number of transmitter cores configurable so that the attacker can adapt them to the expected system utilization.

Turbo Boost regards cores as active whenever they are not in a deep sleep state, i.e., not in a C-state higher than C3~\cite{intel2008turboboostwhitepaper}.
The easiest method to enforce a core to be active is, as described by Alagappan et al.~\cite{alagappan_dfs_2017}, to keep the core busy executing code.
An attacker can implement this method purely in user space without the need for elevated privileges.
Our prototype repeatedly executes long sequences of \texttt{NOP} instructions\footnote{In general, the choice of instructions does not have an impact on the frequency of other cores -- even the frequency reduction caused by AVX2 and AVX-512 instructions only affects a single core.} to keep the core from sleeping. 

Conversely, to transmit a zero, the core has to go to sleep to allow other cores to raise the turbo frequency to the next level.
On Intel CPUs, sleep states are entered by the core when the operating system executes the privileged \texttt{MWAIT} instruction with the appropriate flags set~\cite[Vol. 2B 4-159]{intel_intel_2018}.
User space programs are therefore not able to directly halt a core by themselves.
However, both for power and performance reasons, modern operating systems will automatically put idle cores to sleep.
Therefore, our prototype makes sure that all transmitter processes either execute \texttt{usleep()} or \texttt{sem\_wait()} during \enquote{0}-bits so that the corresponding cores -- if the operating system does not schedule any other program on them -- idle and are sent to sleep.

As described above, the activation and deactivation of multiple cores happens synchronously.
Our test has shown that spawning new processes each time to generate additional CPU load induces significant delay, most likely due to the overhead of forking a new process. 
Therefore, our prototype creates a set of child processes at startup and uses a semaphore to wake them up as required.

\subsection{Receiver}
\label{sec:impl:receiver}

\begin{listing}
	\begin{lstlisting}[frame=single,language=C,xleftmargin=15pt,xrightmargin=5pt]
start = RDTSC;
counter = 0;
while(RDTSC - start < time_frame) {
	counter++;
}
	\end{lstlisting}
	\caption{To measure the frequency, the receiver measures the amount of work performed in a fixed amount of time (pseudocode).}
	\label{fig:implementation:receiver}
\end{listing}

As the transmitter modulates the information onto the turbo frequency, the receiver needs to be able to determine the current turbo frequency of its core.
Although the receiver can read the frequency through the sysfs file system~\cite{alagappan_dfs_2017}, doing so causes a lot of overhead and can easily be detected and blocked by the operating system.
Instead, to be independent of the operating system and have a higher signal detecting frequency, we chose to measure the current frequency with the loop shown in Listing~\ref{fig:implementation:receiver}. 
This loop counts the operations executed in a fixed amount of time which are proportional to the current frequency.
Note that the current frequency of a core does not need to be a turbo frequency if the operating system decides that the utilization of the core is low enough that power-saving reduced frequencies are preferable.
However, the loop generates enough load for the receiver's core to use the turbo frequency.

Listing~\ref{fig:implementation:receiver} uses the \texttt{RDTSC} instruction to measure time.
The \texttt{RDTSC} instruction returns the core's time stamp counter (TSC) which on recent CPUs is independent of the current core clock speed~\cite[Vol. 3B 17.17]{intel_intel_2018}.
On modern out-of-order CPUs, however, the resulting time stamp often does not correlate to a single point in the program, as these CPUs reorder other instructions around the \texttt{RDTSC} instruction.
Such reordering could therefore cause wrong measurements of the runtime of the timing loop.
\texttt{RDTSCP} provides the functionality of \texttt{RDTSC}, but waits until all previous instructions have executed~\cite[Vol. 2B 4-544]{intel_intel_2018}.
However, using \texttt{RDTSCP} has the disadvantage that some hypervisors, such as KVM, disallow the use of this instruction\cite{gulmezoglu_faster_2015}.
Therefore, our prototype uses the \texttt{LFENCE} instruction that serializes all load-from-memory instructions and thereby prevents most reordering.

The time stamp counter itself is not only precise, as required for high bit rates, but also provided directly by the CPU and therefore largely independent of the operating system.
Even though the time presented by the operating system would be precise enough, the receiver would have to rely on the operating system to provide it with the correct time.
The operating system could also potentially disable the use of the \texttt{RDTSC} instruction, but the instruction is widely used to measure exact time.
In particular, the instruction is used on Linux systems to implement \texttt{clock\_gettime()} and \texttt{gettimeofday()} without any system call\footnote{Linux 5.2, \code{arch/x86/entry/vdso/vclock\_gettime.c}, l. 127ff}, so blocking \texttt{RDTSC} would reduce performance.
Even if this instruction is blocked, though, the receiver can either use the normal system clock at the cost of reduced throughput.
Alternatively, it might even be possible to generate a high resolution clock by executing code whose runtime is not proportional to the CPU frequency (e.g., due to cache-misses), similar to the techniques described by Kohlbrenner et al.~\cite{kohlbrenner2016timer}.

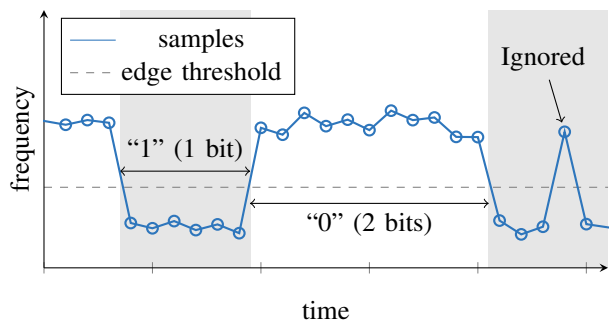
\begin{figure}
	\begin{tikzpicture}
	\begin{axis}[
	ylabel=frequency,
	xlabel=time,
	axis x line = bottom,axis y line = left,
	xticklabels={,,},
	ymajorticks=false,
	ymin=0,
	ymax=1.6,
	xmin=0,
	xmax=5.2,
	width=9cm,
	height=5cm,
	legend entries={
		samples,
		edge threshold,
	},
	legend pos=north west,
	axis background/.style={%
		preaction={
			path picture={
				\coordinate (e1bottom) at (axis cs:0.7,0);
				\coordinate (e2top) at (axis cs:1.9,1.7);
				\coordinate (e3bottom) at (axis cs:4.1,0);
				\coordinate (righttop) at (axis cs:5.2,1.7);
				\fill [gray!20] (e1bottom) rectangle (e2top);
				\fill [gray!20] (e3bottom) rectangle (righttop);
				\coordinate (tleft) at (axis cs:0,0.5);
				\coordinate (tright) at (axis cs:5.2,0.5);
				\draw [gray,sharp plot,dashed] (tleft) -- (tright);
				\coordinate (a1left) at (axis cs:0.7,0.6);
				\coordinate (a1right) at (axis cs:1.9,0.6);
				\draw[<->] (a1left) to node [above, pos=0.5] () {\enquote{1} (1 bit)} (a1right);
				\coordinate (a2left) at (axis cs:1.9,0.4);
				\coordinate (a2right) at (axis cs:4.1,0.4);
				\draw[<->] (a2left) to node [below, pos=0.5] () {\enquote{0} (2 bits)} (a2right);
	}}},
	]
	\addlegendimage{kitblue}
	\addlegendimage{dashed,gray}
	
	\addplot[mark=o, thick, color=kitblue] coordinates {
		(-0.01, 0.91284746)
		(0.2, 0.88799286)
		(0.4, 0.9170874)
		(0.6, 0.9003488)
		(0.8, 0.27817914)
		(1, 0.24554078)
		(1.2, 0.29017842)
		(1.4, 0.23427413)
		(1.6, 0.26960695)
		(1.8, 0.21512863)
		(2, 0.86903536)
		(2.2, 0.82626176)
		(2.4, 0.9618386)
		(2.6, 0.8792818)
		(2.8, 0.9197516)
		(3, 0.8546427)
		(3.2, 0.9755878)
		(3.4, 0.9167617)
		(3.6, 0.9332546)
		(3.8, 0.81277233)
		(4, 0.81133455)
		(4.2, 0.2934058)
		(4.4, 0.20772018)
		(4.6, 0.25506788)
		(4.8, 0.84416906)
		(5, 0.2712007)
		(5.21, 0.25009453)
	};
	\node (label) at (4.6, 1.3) {Ignored};
	\node (ignored) at (4.8, 0.84416906) {};
	\draw[->] (label) -- (ignored);
	\end{axis}
	\end{tikzpicture}
	\caption{
		To recover the data from the signal, the receiver detects edges and measures the time between edges to determine the number of bits.
		Outliers consisting of one or two samples are ignored.
	}
	\label{fig:implementation:edgedetection}
\end{figure}

As the receiver is not perfectly synchronized with the transmitter, a single frequency measurement per transmitted bit is not sufficient.
Our prototype samples the CPU frequency at a significantly higher rate than the bit rate of the signal and then uses edge detection to recover the signal.
As shown in Figure~\ref{fig:implementation:edgedetection}, edge detection only detects changes in the signal from \enquote{1} to \enquote{0} and vice versa, but not the number of identical bits in-between.
Therefore, our prototype measures the time between the two edges and divides it by the time per bit in order to get the transmitted data.
The edge detection categorizes samples depending on whether they are below or above a threshold.
In our prototype, the threshold value is determined manually in order to increase reproducibility of the results, although, as the turbo frequency levels are discrete, automatic calculation of the threshold is possible.

The main problem of simple edge detection is that during frequency sampling each single outlier immediately is recognized as an edge and is therefore detected as a new bit.
Such errors can be caused by, for example, concurrent signal recovery, interrupt handling on the receiver's core, or migration between cores.
As these errors are fairly frequent and each error potentially causes the receiver to lose synchronization to the sender, the edge detection algorithm discards changes of the signal which are significantly shorter than a single bit.

\subsection {Communication Protocol}
\label{sec:imp:communication_protocol}

\begin{figure}
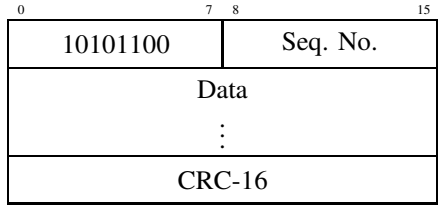

	\centering
	\begin{bytefield}[bitwidth=1em]{16}
		\bitheader{0,7,8,15} \\
		\bitbox{8}{10101100} & 	\bitbox{8}{Seq. No.}\\
		\wordbox[lr]{1}{Data} \\
		\wordbox[lr]{1}{$\vdots$ \\[1ex]} \\
		\bitbox{16}{CRC-16} 
	\end{bytefield}
	\caption{Each packet consists of the synchronization sequence, a sequence number, the data and a checksum.}
	\label{fig:design:packet}
\end{figure}

In addition to the transmission mechanism of the receiver and transmitter, we added a simple communication protocol which has to fulfill mainly two tasks.
First, the mechanism described above requires sender and receiver to be synchronized so that the receiver recognizes when the sender starts transmitting information.
Second, even though the most common intermittent transmission errors are covered by our edge detection mechanism, other errors might still require retransmitting information that was lost to ensure robust error-free communication.
The communication protocol enables the transmitter to determine whether packets were lost or corrupted and have to be retransmitted..
Figure~\ref{fig:design:packet} shows the structure of a single data packet which contains fields to deal with both these problems.

As our communication channel does not have any external synchronization signal, our prototype implements synchronization between sender and receiver via the transmission of a special bit sequence at the beginning of every transmission, similar to POWERT~\cite{khatamifard_powert_2019}.
Whenever the receiver recognizes the corresponding characteristic frequency changes, it starts recording incoming data.
The bit sequence has to be long and complex enough that it is unlikely to occur by accident when the scheduler lets other processes run.
Short sequences, like \texttt{1100}, are too likely to occur and therefore we chose the longer and more complex sequence \texttt{10101100} for the synchronization.
Note that occasional misinterpretation of background noise as the start of the transmission by the receiver is not particularly problematic as the received data will be recognized as being erroneous and will be discarded as described below.

Even if the synchronization header is correctly recognized, we still have to handle other errors during the transmission of the data.
Similar to most network protocols, our approach uses checksums and a retransmission mechanism to handle errors in the received data.
We add a CRC-16 checksum to the transmitted data in order to detect errors.

To notify the sender about a failed transmission, the receiver checks the checksum and replies with a short acknowledgment message back to the original sender.
If the sender does not receive that message in time or if the acknowledgment itself is corrupted, the sender retransmits the data.
During this process, sender and receiver are switching their role to allow the acknowledgment to be send back to the original sender. 
For the acknowledgement packets, while the underlying transmission mechanism is fully bidirectional, we use a simplified format without the possibility to send any additional data back to the original transmitter.
The packet format of our covert channel could be trivially extended to allow full bidirectional communication, albeit at slightly higher overhead.

In order to restrict the data that has to be retransmitted after each error, the sender splits the payload into multiple packets of fixed size.
Each packet contains the synchronization bit sequence, a sequence number to identify the packet, the fixed size data and the checksum over the rest of the packet.
Similar to reliable network protocols such as TCP, each acknowledgment packet contains the sequence number of the last correctly received packet.
The sender only starts transmitting the next packet once a correct acknowledgment for the previous packet has been received.

Note that, due to the low throughput of our covert channel, the packets are of a fixed size and do not contain any additional information about the length of the data carried.
It is left to the application to implement correct padding.
Our packets each contain 8 bytes of data as our experiments have shown this size to be a good compromise between the overhead per transmitted byte and the amount of retransmitted data. 

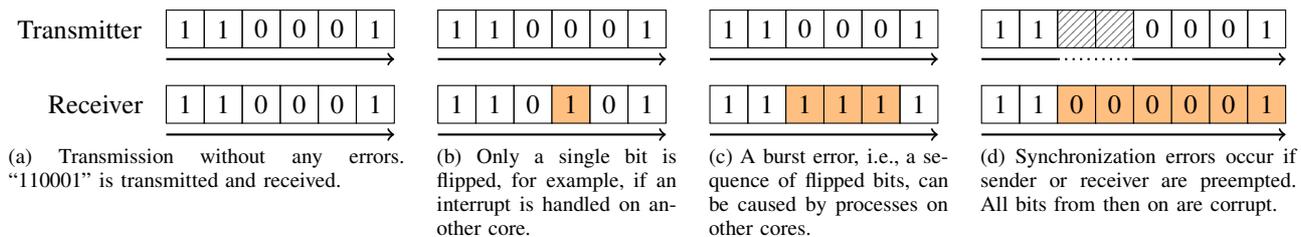
\begin{figure*}[t]
	\centering
	\subfloat[Transmission without any errors. \enquote{110001} is transmitted and received.\label{fig:error:normal}]{%
		\begin{tikzpicture}[]

\coordinate (A) at (0,1);
\node (sender) [left of=A,yshift=7,xshift=-4] {Transmitter};

\draw[draw=black] (A) rectangle ++(0.5,0.5) node[pos=.5] {1};

\coordinate[shift={(0.5,0)}] (B) at (A);
\draw[draw=black] (B) rectangle ++(0.5,0.5) node[pos=.5] {1};

\coordinate[shift={(0.5,0)}] (C) at (B);
\draw[draw=black] (C) rectangle ++(0.5,0.5) node[pos=.5] {0};

\coordinate[shift={(0.5,0)}] (D) at (C);
\draw[draw=black] (D) rectangle ++(0.5,0.5) node[pos=.5] {0};

\coordinate[shift={(0.5,0)}] (E) at (D);
\draw[draw=black] (E) rectangle ++(0.5,0.5) node[pos=.5] {0};

\coordinate[shift={(0.5,0)}] (F) at (E);
\draw[draw=black] (F) rectangle ++(0.5,0.5) node[pos=.5] {1};

\draw[->,thick] (0,0.85) to (3.0,0.85);

\coordinate (G) at (0,0);
\node (sender) [left of=G,yshift=7,xshift=2] {Receiver};

\draw[draw=black] (G) rectangle ++(0.5,0.5) node[pos=.5] {1};
\coordinate[shift={(0.5,0)}] (H) at (G);
\draw[draw=black] (H) rectangle ++(0.5,0.5) node[pos=.5] {1};

\coordinate[shift={(0.5,0)}] (I) at (H);
\draw[draw=black] (I) rectangle ++(0.5,0.5) node[pos=.5] {0};

\coordinate[shift={(0.5,0)}] (J) at (I);
\draw[draw=black] (J) rectangle ++(0.5,0.5) node[pos=.5] {0};

\coordinate[shift={(0.5,0)}] (K) at (J);
\draw[draw=black] (K) rectangle ++(0.5,0.5) node[pos=.5] {0};

\coordinate[shift={(0.5,0)}] (L) at (K);
\draw[draw=black] (L) rectangle ++(0.5,0.5) node[pos=.5] {1};

\draw[->,thick] (0,-0.15) to (3.0,-0.15);

\end{tikzpicture}}
	\hspace{0.3cm}
	\subfloat[Only a single bit is flipped, for example, if an interrupt is handled on another core.\label{fig:error:singlebit}]{%
		\begin{tikzpicture}[]

\coordinate (A) at (0,1);

\draw[draw=black] (A) rectangle ++(0.5,0.5) node[pos=.5] {1};

\coordinate[shift={(0.5,0)}] (B) at (A);
\draw[draw=black] (B) rectangle ++(0.5,0.5) node[pos=.5] {1};

\coordinate[shift={(0.5,0)}] (C) at (B);
\draw[draw=black] (C) rectangle ++(0.5,0.5) node[pos=.5] {0};

\coordinate[shift={(0.5,0)}] (D) at (C);
\draw[draw=black] (D) rectangle ++(0.5,0.5) node[pos=.5] {0};

\coordinate[shift={(0.5,0)}] (E) at (D);
\draw[draw=black] (E) rectangle ++(0.5,0.5) node[pos=.5] {0};

\coordinate[shift={(0.5,0)}] (F) at (E);
\draw[draw=black] (F) rectangle ++(0.5,0.5) node[pos=.5] {1};

\draw[->,thick] (0,0.85) to (3.0,0.85);

\coordinate (G) at (0,0);

\draw[draw=black] (G) rectangle ++(0.5,0.5) node[pos=.5] {1};
\coordinate[shift={(0.5,0)}] (H) at (G);
\draw[draw=black] (H) rectangle ++(0.5,0.5) node[pos=.5] {1};

\coordinate[shift={(0.5,0)}] (I) at (H);
\draw[draw=black] (I) rectangle ++(0.5,0.5) node[pos=.5] {0};

\coordinate[shift={(0.5,0)}] (J) at (I);
\draw[draw=black,fill=orange!50] (J) rectangle ++(0.5,0.5) node[pos=.5] {1};

\coordinate[shift={(0.5,0)}] (K) at (J);
\draw[draw=black] (K) rectangle ++(0.5,0.5) node[pos=.5] {0};

\coordinate[shift={(0.5,0)}] (L) at (K);
\draw[draw=black] (L) rectangle ++(0.5,0.5) node[pos=.5] {1};

\draw[->,thick] (0,-0.15) to (3.0,-0.15);

\end{tikzpicture}}
	\hspace{0.3cm}
	\subfloat[A burst error, i.e., a sequence of flipped bits, can be caused by processes on other cores.\label{fig:error:burst}]{%
		\begin{tikzpicture}[]

\coordinate (A) at (0,1);

\draw[draw=black] (A) rectangle ++(0.5,0.5) node[pos=.5] {1};

\coordinate[shift={(0.5,0)}] (B) at (A);
\draw[draw=black] (B) rectangle ++(0.5,0.5) node[pos=.5] {1};

\coordinate[shift={(0.5,0)}] (C) at (B);
\draw[draw=black] (C) rectangle ++(0.5,0.5) node[pos=.5] {0};

\coordinate[shift={(0.5,0)}] (D) at (C);
\draw[draw=black] (D) rectangle ++(0.5,0.5) node[pos=.5] {0};

\coordinate[shift={(0.5,0)}] (E) at (D);
\draw[draw=black] (E) rectangle ++(0.5,0.5) node[pos=.5] {0};

\coordinate[shift={(0.5,0)}] (F) at (E);
\draw[draw=black] (F) rectangle ++(0.5,0.5) node[pos=.5] {1};

\draw[->,thick] (0,0.85) to (3.0,0.85);

\coordinate (G) at (0,0);

\draw[draw=black] (G) rectangle ++(0.5,0.5) node[pos=.5] {1};
\coordinate[shift={(0.5,0)}] (H) at (G);
\draw[draw=black] (H) rectangle ++(0.5,0.5) node[pos=.5] {1};

\coordinate[shift={(0.5,0)}] (I) at (H);
\draw[draw=black,fill=orange!50] (I) rectangle ++(0.5,0.5) node[pos=.5] {1};

\coordinate[shift={(0.5,0)}] (J) at (I);
\draw[draw=black,fill=orange!50] (J) rectangle ++(0.5,0.5) node[pos=.5] {1};

\coordinate[shift={(0.5,0)}] (K) at (J);
\draw[draw=black,fill=orange!50] (K) rectangle ++(0.5,0.5) node[pos=.5] {1};

\coordinate[shift={(0.5,0)}] (L) at (K);
\draw[draw=black] (L) rectangle ++(0.5,0.5) node[pos=.5] {1};

\draw[->,thick] (0,-0.15) to (3.0,-0.15);

\end{tikzpicture}}
	\hspace{0.3cm}
	\subfloat[Synchronization errors occur if sender or receiver are preempted. All bits from then on are corrupt.\label{fig:error:sync}]{%
		\begin{tikzpicture}[]

\coordinate (A) at (0,1);

\draw[draw=black] (A) rectangle ++(0.5,0.5) node[pos=.5] {1};

\coordinate[shift={(0.5,0)}] (B) at (A);
\draw[draw=black] (B) rectangle ++(0.5,0.5) node[pos=.5] {1};

\coordinate[shift={(0.5,0)}] (X) at (B);
\draw[draw=black,pattern=north east lines, pattern color=gray] (X) rectangle ++(0.5,0.5) node[pos=.5] {};

\coordinate[shift={(0.5,0)}] (Y) at (X);
\draw[draw=black,pattern=north east lines, pattern color=gray] (Y) rectangle ++(0.5,0.5) node[pos=.5] {};

\coordinate[shift={(0.5,0)}] (C) at (Y);
\draw[draw=black] (C) rectangle ++(0.5,0.5) node[pos=.5] {0};

\coordinate[shift={(0.5,0)}] (D) at (C);
\draw[draw=black] (D) rectangle ++(0.5,0.5) node[pos=.5] {0};

\coordinate[shift={(0.5,0)}] (E) at (D);
\draw[draw=black] (E) rectangle ++(0.5,0.5) node[pos=.5] {0};

\coordinate[shift={(0.5,0)}] (F) at (E);
\draw[draw=black] (F) rectangle ++(0.5,0.5) node[pos=.5] {1};

\draw[-,thick] (0,0.85) to (1,0.85);
\draw[-,dotted,thick] (1,0.85) to (2,0.85);
\draw[->,thick] (2,0.85) to (4,0.85);

\coordinate (G) at (0,0);

\draw[draw=black] (G) rectangle ++(0.5,0.5) node[pos=.5] {1};
\coordinate[shift={(0.5,0)}] (H) at (G);
\draw[draw=black] (H) rectangle ++(0.5,0.5) node[pos=.5] {1};

\coordinate[shift={(0.5,0)}] (I) at (H);
\draw[draw=black,fill=orange!50] (I) rectangle ++(0.5,0.5) node[pos=.5] {0};

\coordinate[shift={(0.5,0)}] (J) at (I);
\draw[draw=black,fill=orange!50] (J) rectangle ++(0.5,0.5) node[pos=.5] {0};

\coordinate[shift={(0.5,0)}] (K) at (J);
\draw[draw=black,fill=orange!50] (K) rectangle ++(0.5,0.5) node[pos=.5] {0};

\coordinate[shift={(0.5,0)}] (L) at (K);
\draw[draw=black,fill=orange!50] (L) rectangle ++(0.5,0.5) node[pos=.5] {0};

\coordinate[shift={(0.5,0)}] (M) at (L);
\draw[draw=black,fill=orange!50] (M) rectangle ++(0.5,0.5) node[pos=.5] {0};

\coordinate[shift={(0.5,0)}] (N) at (M);
\draw[draw=black,fill=orange!50] (N) rectangle ++(0.5,0.5) node[pos=.5] {1};

\draw[->,thick] (0,-0.15) to (4,-0.15);

\end{tikzpicture}}
	\caption{Possible transmission errors in our side channel -- our system suffers from the similar types of errors as cache-based covert channels~\cite{maurice_hello_2017}, except that we expect burst errors to be more frequent.}
	\label{fig:error}
\end{figure*}

\subsection{Error Correction}
\label{sec:impl:error_correction}

Although the retransmissions of corrupt packets are sufficient to achieve a robust communication channel, simple error detection might not be the most efficient method to handle transmission errors.
Especially for higher error rates, error correction codes might be able to repair some corrupt data, might reduce the number of retransmissions, and might thereby increase performance.
As, however, the choice of error handling mechanisms depends on the type and rate of errors, in the following, we first discuss the different error types and then show why adding further error correction is disadvantageous for our covert channel.

Our covert channel is supposed to be usable on a normal system with all its background tasks which disturb the transmitted signal.
As shown in Figure \ref{fig:error}, the transmissions in our covert channel are affected by the same types of errors as in the cache-based covert channel described by Maurice et al.~\cite{maurice_hello_2017}. 
Single bit flip and burst errors are caused by the scheduler of the operating system which chooses processes to run on sleeping cores, which causes the cores to wake up and have an impact on the turbo frequency. 
We counted the frequency changes caused by other cores becoming active during one second and measured their length.
As shown in Table \ref{table:errors}, on an idle system most of these frequency changes are short and will not cause transmission errors if the time per transmitted bit is at least several milliseconds.
Therefore, reducing the transmission rate to reduce transmission errors provides a simple approach to limit the number or required retransmissions.
It presents a trade-off between the number of retransmissions and the maximum achievable throughput though.
Also, even arbitrarily low bit rates do not fully guarantee that no errors occur as there is no bound on the run time of other processes, so some retransmissions might be required.

As error correction codes are able to repair corrupt data at the receiving end of the channel, they have the potential to reduce the number of retransmissions which can have a positive impact on throughput.
However, the codes itself need to be appended to each packet and cause overhead.
Therefore, the use of error correction codes also presents a trade-off where the utility depends on the transmission error rate and the type of the errors.
To determine whether error correction is advantageous in our setup, we constructed a simplified version of our covert channel without retransmissions and acknowledgement packets and we recorded the packet content at the receiver on an idle system.
For each packet, we used the data received to determine whether an error-correcting code could have successfully repaired the packet.
For our analysis, we chose the widely used Reed-Solomon code with 4 bytes of parity which is able to recover up to two bytes of data.
Using more parity bytes would allow recovering more bytes of data but would increase the packet transmission time.
In any case, the Reed-Solomon does not remove the need for the CRC-16 checksum, as the retransmission mechanism would still be required for robustness.
To simplify estimating the transmission rate, we assume that a retransmission of a packet will be without errors.

The measured number of corrupted bytes per transmission shows that most errors could have been corrected at all transmission rates where the covert channel is viable.
At a high transmission rate of only \SI{5}{\ms} per bit, 29 out of 40 packets were without any errors and out of these 11 broken packets 7 could have been fixed by the Reed-Solomon code.
With retransmissions as well as the use of the Reed-Solomon code, we calculate a theoretical data rate of \SI{107}{\bps}.
Without Reed-Solomon we expect a data rate of approximately \SI{121}{\bps}.
This analysis shows that, despite the correctable errors, the overhead of the Reed-Solomon code would be enough to offset the advantages.

With only four broken transmissions, our experiment showed the lowest error rate at a data rate of \SI{10}{\ms} per bit.
All of these four corrupted transmissions could have been fixed by the Reed-Solomon code.
However, even in this situation, error correction would not yield a net advantage.
With the Reed-Solomon code, we calculate a throughput of \SI{59}{\bps}, whereas we expect the system to achieve \SI{70}{\bps} without error correction.
As none of these setups promise any advantage due to error correction, our prototype only employs error detection, retransmitting any corrupted packet as necessary.

\begin{table}
	\caption{Number of involuntary frequency changes during one second along with their respective duration on an idle system with only background services running}
	\label{table:errors}
	\begin{center}
		\begin{tabular}{cc}
			\toprule
			Duration & Involuntary Frequency Changes \\
			\midrule
			1 \si{\milli\second} & 109 \\
			2 \si{\milli\second} & 5 \\
			3 \si{\milli\second} & 2 \\
			4 \si{\milli\second} & 1 \\
			6 \si{\milli\second} & 1 \\
			\bottomrule
		\end{tabular}
	\end{center}

\end{table}

\section{Evaluation}
\label{sec:evaluation}

We conducted an evaluation of our implementation consisting of two parts:
First, in Section~\ref{eval:throughput}, we measure the throughput of our prototype on an idle system for different bit rates and show how, despite the runtime overhead added by the retransmission protocol, our prototype achieves a throughput close to the prediction by Khatamifard et al.~\cite{khatamifard_powert_2019}.
Second, in Section~\ref{sec:eval:backgroundload}, we show that our covert channel is able to achieve sufficient throughput even in a system with significant background load.
To finalize our evaluation, we show in Section~\ref{sec:vm} that our covert channel is even able to communicate between two VMs running on the same CPU. 

\subsection{Setup}
\label{eval:setup}
As our covert channel is usable in both server and desktop environments, we chose to use the Intel Xeon Silver 4108 CPU as our primary target system.
Although this specific CPU is usually found in servers, most recent desktop CPUs by Intel have a similar architecture with similar power management.
This CPU is based on the Skylake architecture and has 8 physical cores with a base frequency of \SI{1.8}{\giga\hertz} and a max turbo frequency of \SI{3.0}{\giga\hertz} with Turbo Boost 2.0. 
The system runs Fedora 28 and is sufficiently cooled so that our experiments do not trigger thermal throttling.
For the experiments in Section~\ref{eval:throughput}, the system is idle during the experiments, i.e., only the terminal and the background tasks of the operating system are intermittently active.

All experiments are executed with deactivated hyper-threading to prevent running the transmitter and the receiver on the same physical core, as our covert channel was only designed for inter-core communication.
If transmitter and receiver are on the same physical core and share functional units, other covert channels such as the one described by Wang et al.~\cite{wang_covert_2006} are available instead.
It is likely that the two approaches can be combined to create a covert channel that works both for communication between the hyper threads of a single core and between different physical cores.
For example, sender and receiver could alternate between the different methods to send messages until one of them succeeds.
We expect such an approach to become less important, though, as recently discovered high-risk side channels that require hyper-threading~\cite{schwarz2019zombieload,aldaya2019port} likely trigger increasing deployment of counter measures.
For example, hyper-threading can be completely deactivated~\cite{marshall2010msAzure} or in order to reduce the performance impact the OS can instead use scheduling algorithms that prevent applications from different sources to execute on the same physical core~\cite{corbet_many_2019}.

\subsection{Idle Target System}
\label{eval:throughput}

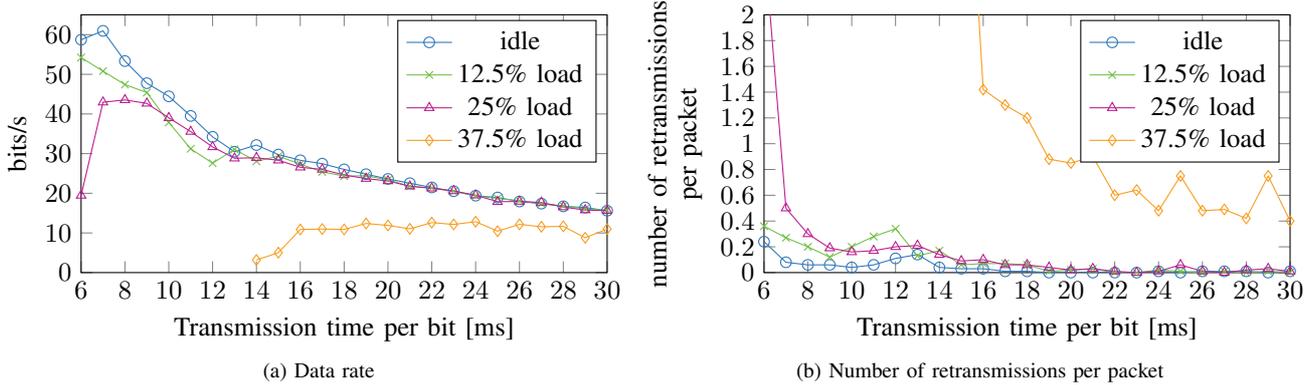
\begin{figure*}[t]
	\centering
	\subfloat[Data rate\label{fig:eval:multi_thread:transmissionrate}]{%
	\centering
	\begin{tikzpicture}
	\begin{axis}[
	xlabel={Transmission time per bit [ms]},
	ylabel=bits/s,
	width=8.5cm,
	height=5cm,
	xmin=6,
	xmax=30,
	ymin=0,
	ymax=65,
	xtick distance=2,
	ytick distance=10
	]
	\pgfplotstableread{data/multi_core.txt}\mydata;
	\addplot[color=kitblue, mark=o] table [x=t, y=datarate] {\mydata};
	\pgfplotstableread{data/multi_core_n1.txt}\mydata;
	\addplot[color=kitgreen, mark=x] table [x=t, y=datarate] {\mydata};
	\pgfplotstableread{data/multi_core_n2.txt}\mydata;
	\addplot[color=kitpurple, mark=triangle] table [x=t, y=datarate] {\mydata};
	\pgfplotstableread{data/multi_core_n3.txt}\mydata;
	\addplot[color=kitorange, mark=diamond] table [x=t, y=datarate] {\mydata};
	\legend{idle,12.5\% load, 25\% load, 37.5\% load}
	\end{axis}
	\end{tikzpicture}
}
\subfloat[Number of retransmissions per packet\label{fig:eval:multi_thread:retransmitts}]{%
	\centering
	\begin{tikzpicture}
	\begin{axis}[
	ylabel style={align=center},
	xlabel={Transmission time per bit [ms]},
	ylabel={number of retransmissions\\{}per packet},
	width=8.5cm,
	height=5cm,
	xmin=6,
	xmax=30,
	ymin=0,
	ymax=2,
	xtick distance=2,
	ytick distance=0.2
	]
	\pgfplotstableread{data/multi_core.txt}\mydata;
	\addplot[color=kitblue, mark=o] table [x=t, y=retransmission] {\mydata};
	\pgfplotstableread{data/multi_core_n1.txt}\mydata;
	\addplot[color=kitgreen, mark=x] table [x=t, y=retransmission] {\mydata};
	\pgfplotstableread{data/multi_core_n2.txt}\mydata;
	\addplot[color=kitpurple, mark=triangle] table [x=t, y=retransmission] {\mydata};
	\pgfplotstableread{data/multi_core_n3.txt}\mydata;
	\addplot[color=kitorange, mark=diamond] table [x=t, y=retransmission] {\mydata};
	\legend{idle,12.5\% load, 25\% load, 37.5\% load}
	\end{axis}
	\end{tikzpicture}
}
	\caption{
		Data rate and average number of retransmissions required for a transmission of 80~bytes of data with different background loads. With increasing background load, the number of transmission errors rises which causes more retransmissions and therefore decreases the data rate. 
		For example, at 25\% background load and 6 milliseconds transmission time per bit, the number of retransmissions per packet is at 2.58, which means that every packet is sent about two and a half times before it was delivered successful.
	}
	\label{fig:eval:multiple_thread_transmission}
\end{figure*}

The main goal of our prototype is to provide a practical covert channel that, in contrast to other approaches, is optimized for robustness in the face of significant background CPU load and other sources for transmission errors.
The communication protocol required, however, potentially introduces significant overhead.
In this section, we show that the solution is still able to provide throughput comparable to other approaches.

The throughput of a covert channel such as ours depends on several variables, amongst them the overhead of the communication protocol, the raw transmission time per bit, and the number of retransmissions caused by transmission errors.
The communication protocol is fixed, but the bit rate is variable and the number of transmission errors depends on the bit rate as well as the background load on the system.
To determine the achievable throughput at a specific bit rate, we use the covert channel to transmit 80 bytes of data while we measure the time between the start of the transmission and the reception of the last acknowledgement packet.
We repeat this experiment for different bit rates to determine the optimal throughput.

The results for an idle system are shown as the blue line in Figure~\ref{fig:eval:multi_thread:transmissionrate}.
As transmission errors have an unpredictable influence on the throughput, we repeated each test 10 times and averaged the results.
The highest throughput of \SI{61}{\bps} is reached at a raw transmission time of \SI{7}{\milli\second} per bit.
Intuitively, the throughput increases with increasing transmission speed.
Below \SI{7}{\milli\second} per bit, the throughput decreases again, as the signal conditioning of our prototype -- which is supposed to filter out short frequency changes caused by interrupts on other cores -- fails to detect the short frequency changes caused by the transmitter.
This effect can be seen as an increase in Figure~\ref{fig:eval:multi_thread:retransmitts}, which shows the number of retransmissions per packet during the experiment.
For a similar system, Khatamifard et al.~\cite{khatamifard_powert_2019} have determined the maximum throughput to be approximately \SI{120}{\bps} -- our system has significantly lower throughput due to the communication protocol, it would be the same if we removed our error detection and communication protocol.
Note that further optimization -- e.g., letting the transmitter send multiple packets in one batch and then letting the receiver send an accumulative acknowledgement -- might improve the throughput of our prototype.

\subsection{Background Load}
\label{sec:eval:backgroundload}

To demonstrate that our system is able to cope with significant background load caused by other processes on the same system, we repeat the experiment from the last section.
In parallel to our prototype, we execute an instance of the x265 video encoder to generate background CPU load. 
We vary the number of threads used by the video encoder.
x265 was chosen as a workload as it produces a mostly stable background load with some I/O operations. 
For workloads with higher variation to cause transmission errors, two conditions have to be fulfilled: First, the utilization temporarily needs to be higher than the maximum supported utilization, and second, the utilization spikes need to be long enough to be detected as a \enquote{1}-bit.
Although some cloud workloads violate the latter, others do not.
In addition to the experiments described below, we conducted an additional test with a kernel compilation benchmark, where \texttt{make} spawns processes on varying cores and I/O interrupts are often triggered on additional cores, yet our prototype was able to transmit data.

Again, Figure~\ref{fig:eval:multiple_thread_transmission} shows the results of these experiments.
The figure shows that the achievable throughput varies with the utilization of the system:
As shown, our covert channel is able to cope well with up to 25\% background load. 
Except at the higher bit rates, the throughput is nearly the same as without background load. 
At higher bit rates, spikes in the CPU load cause numerous transmission errors.

With 37.5\% background load, our covert channel shows a significant lower data rate. 
Note that this background load does not include the processes used by the sender and receiver to implement the covert channel. 
The resulting utilization as seen by the operating system is therefore higher.
At 37.5\% background load, on average three cores are fully utilized by the background processes, as well as one by the receiver. 
As shown in Table~\ref{tab:turbo_frequenzen}, the CPU selects the lowest turbo frequency whenever more then 4 cores are active.\footnote{Note that these values are for the 8-core system on which we developed our prototype. The number of turbo frequencies is higher for systems with larger core counts, so the prototype is able to work in the presence of higher background utilization on such (likely more representative) systems.}
Therefore, any additional core will cause the frequency drop. 
As we do not prohibit any background services from executing and as the background load is not uniform, load fluctuations cause frequent transmission errors.
Nevertheless, the covert channel is still functional despite the significant amount of background load.

Even cloud servers, where cloud providers try to increase utilization by placing workloads from different users on one system, are frequently operated at such utilization levels.
For example, Garraghan~\cite{garraghan_analysis_2013} quote an average utilization from 28.34\% to 55.66\% for the Google cloud data center. 
In our case, the number of cores used by background tasks only needs to be lower than the number of active cores of the second to last turbo frequency step, minus the one core needed for the receiver. 
In addition, the operator of a system under attack not only sees the background load but also the load caused by the covert channel itself.
The sum of both is significantly higher than the background load itself, thereby potentially deterring the operator of the system from placing more load on it.
Overall, many server systems are therefore likely operated at utilization levels where our covert channel is viable.

Furthermore, many server systems use server CPUs with more cores than the Xeon Silver used in our experiments.
For example, the Xeon Gold 6130 changes its turbo frequency when surpassing 75\% load, yielding a frequency change for each four additional active cores~\cite{xeonscalableupdate}.
On such a system, this additional frequency level can likely be used to transmit information at even higher background load.

\subsection{TurboCC Between VMs}
\label{sec:vm}

\begin{table*}
	\centering
	\begin{minipage}{\textwidth}
		\centering
		\caption{
			Number of rescheduling interrupts during a transmission of 100 bytes between processes and between VMs.
			Rescheduling interrupts trigger preemption and can cause synchronization errors.
			The cores of the receiver's VM experience higher interrupt rates, which is likely why our prototype only achieves lower throughput.
		}
		\begin{tabular}{c|c|c|c|c|c|c|c|c}
			\toprule
			& \multicolumn{8}{c}{interrupts per core [interrupts/s] } \\
			& 0 & 1 & 2 & 3 & 4 & 5 & 6 & 7\\
			\midrule
			
			processes &	84.33 & \cellcolor{kitgreen!25}0.1~\textsuperscript{a}	&\cellcolor{kitbrown!25}0.03~\textsuperscript{c} & 0.13	&0 &\cellcolor{kitblue!25}0.03~\textsuperscript{b} & 0	&	0\\
			\midrule			
			virtual machines & 131.45 & \cellcolor{kitgreen!25}4.26~\textsuperscript{a} & \cellcolor{kitgreen!25}1.02~\textsuperscript{a} & 0 &  \cellcolor{kitpurple!25}4.43~\textsuperscript{d} & \cellcolor{kitpurple!25}4.10~\textsuperscript{d} & 0.05 & 0 \\
			\bottomrule
		\end{tabular}~\\~\\~\\t
		\colorbox{kitgreen!25}{~\textsuperscript{a}~}~Receiver or receiver's VM\hskip 0.3cm \colorbox{kitblue!25}{~\textsuperscript{b}~}~Transmitter\hskip 0.3cm \colorbox{kitbrown!25}{~\textsuperscript{c}~}~Additional active core\hskip 0.3cm \colorbox{kitpurple!25}{~\textsuperscript{d}~}~Transmitter's VM + active core
		
		\label{table:eval:interrupts}
	\end{minipage}
\end{table*}

Our covert channel neither needs any special instruction nor depends on any particular feature of the operating system and therefore is not limited to communication between processes within one operating system.
As an example for other setups, we setup two virtual machines on one host system to show that the channel can also be used to communicate between virtualized systems.
Both VMs use the KVM hypervisor and are allocated two virtual cores each.
These virtual cores are mapped to different physical cores to prevent the VMs from running on the same core and to provide a more realistic scenario. 
The VMs are running the same stock Fedora 28 as the host system.
With this setup and the same transmission time of \SI{7}{\milli\second} as used above, our prototype only reaches approx. \SI{11}{\bps} with approx. 3.2 retransmissions per sent packet.

To identify possible causes for this increased number of retransmissions, we looked at the number of interrupts occurring during a transmission. 
Handling most interrupts takes little time, so the OS only briefly activates cores or briefly suspends processes.
Our signal detection algorithm is able to cope with such short disruptions.
Therefore, we looked at rescheduling interrupts in particular, because these can cause a process to be suspended much longer and will eventually cause synchronization errors. 
Table \ref{table:eval:interrupts} compares the number of rescheduling interrupts per second between two setups with and without virtualization. 
As three of the four cores allocated to the VMs experience approximately four interrupts per second -- significantly higher than in the setup without virtualization -- we expect that the lower data rate between VMs is caused by these rescheduling interrupts. 
At the selected bit rate of \SI{7}{\milli\second} per bit, a transmission of one packet takes \SI{728}{\milli\second}.
On average, the transmitter and the receiver will therefore get preempted roughly three times per packet during a transmission between VMs. 
These interrupts will eventually cause transmission errors and therefore retransmissions.

Note that core 0 has substantially higher rescheduling interrupt rates than any other core. 
As none of our prototype's processes are running on this core, we think that these interrupts are all caused by the operating system managing the system.
Although these interrupts do cause additional noise for our covert channel, the rate does not differ much between the two setups.

\section{Countermeasures}
\label{sec:counter}
As we have shown that an attacker can use our covert channel to bypass access control, for some setups with sensitive information it might make sense to implement countermeasures to prevent such flow of information.
In the following, we discuss a number of potential countermeasures and their respective drawbacks.

\subsection{Restricting Frequency Management}
\label{sec:counter:restrict}

The most obvious countermeasure against frequency-based covert channels is to prevent load-based frequency changes.
In our case, interaction between the frequencies of the different cores could be prevented by completely disabling Turbo Boost.
The drawback, however, will invariably be reduced performance.
On a system with a desktop CPU based on the Nehalem microarchitecture, Charles et al.~\cite{charles_evaluation_2009} measured 6\% lower performance without Turbo Boost.
To show the impact on newer CPUs, we executed the Parsec benchmark suite~\cite{bienia11benchmarking} on a Xeon Silver 4108 server CPU with Turbo Boost enabled as well as disabled and found an average performance reduction of 23\% (between 15\% and 50\% depending on the benchmark).
Note that these numbers include the single-threaded initialization phase of the benchmarks.
Although disabling Turbo Boost has the potential to improve energy efficiency~\cite{charles_evaluation_2009}, the experiments show that disabling Turbo Boost comes at a significant cost to performance.

As an alternative, it is possible to force the CPU to limit its turbo frequency to the lowest turbo frequency.
Our tests have shown that by preventing CPU sleep states higher than C2 the operating system can keep enough cores active in order to prevent higher turbo frequencies.
As the all-core turbo frequency is higher than the base frequency of the system, the resulting performance reduction is lower compared to disabling Turbo Boost.
For the Parsec benchmark suite, we measured an average overhead of only 5\%, which we expect given the parallel nature of the benchmarks.
Single-threaded performance, however, is expected to experience larger performance reduction.
Also, disabling deeper sleep states greatly increases power consumption of idle and partially loaded systems.
In our case, the idle power consumption of the CPU increased by almost 60\% when the C-states of all cores were restricted.

\subsection{Adding Artificial Noise}
\label{sec:counter:noise}

A countermeasure with potentially lower performance impact would be to randomly add more artificial background load to increase the transmission error rate to the point where no useful data transfer is possible anymore.
The operating system could randomly wake additional cores up to mask the effect of the transmitter.
To prevent the attack, the number of additionally active cores needs to be at least as large as the number of cores controlled by the transmitter as described in Section~\ref{sec:impl:transmitter}.
In practice, the transmitter core count is likely limited if the transmitter is within a virtual machine with a limited number of virtual CPUs.
While we still expect measurable energy and performance overhead due to additional cores being active, the approach has less overhead compared to the alternative countermeasures listed above if the transmitter can only control a limited number of cores.

\subsection{Detecting Frequency Covert Channels}
\label{sec:counter:detecting}

As the methods described above all have drawbacks in terms of performance or energy consumption, it would be beneficial to limit their usage to when a covert channel is detected.
For that reason, mechanisms to detect this kind of covert channel are desired.
Wu et al.~\cite{wu_c2detector:_2014} describe a framework for detecting covert channels that is continuously logging, for example, the system load and then tries to detect abnormal behavior.
Such a technique could be applied to our covert channel as well.
As frequency changes occur on the order of milliseconds, a sufficiently high logging rate of the current turbo frequency is required to detect this covert channel. 

\section{AMD Precision Boost}
\label{sec:amd_precision_boost}
The covert channel presented in this work is based on Intel Turbo Boost 2.0.
Recent AMD processors also have a similar mechanism named AMD Precision Boost~\cite{dezso_amds_2019}.
Similar to Turbo Boost, Precision Boost increases the frequency of the CPU above its base frequency, but in a different way.
The first version of Precision Boost was documented to increase the CPU frequency to the maximum turbo frequency when two cores or less were utilized.
If more than two cores are active, the turbo frequency is set to a far lower turbo frequency. 
Precision Boost 2, in addition, increased the frequency for more than two active cores even further, depending on the number of active cores and as long as the thermal headroom allows it~\cite{cutress2018ryzen2}.
Due to the similar behavior, a similar covert channel can be implemented the same way on AMD CPUs as on Intel ones.
However, our experiments showed that the reaction to load changes is very slow.
Depending on the number of active cores, AMD Zen+ cores implementing Precision Boost 2 can take up to almost \SI{400}{\milli\second} before a core returns to the maximum frequency after other cores go to sleep.
As expected from these delays, our experiments show that our covert channel is only able to reach about \SI{1.08}{\bps} on an idle system with an AMD Ryzen 2700X CPU.
As increased CPU load further reduces the throughput, we do not deem the covert channel to be viable on this specific CPU.
Note that cores from the newer Zen 2 architecture show faster reaction to load changes.
More work has to be conducted to determine whether the covert channel is able to provide satisfactory throughput on these CPUs.

\section{Discussion}
\label{sec:discussion}

Our evaluation has shown that the covert channel can be used to transmit information between two cores, and the throughput is in the same order of magnitude as for other approaches under similar conditions~\cite{alagappan_dfs_2017,miedl_frequency_2018,khatamifard_powert_2019}.
Although a throughput of \SI{61}{\bps} might seem low, an attacker is often only interested in small cryptographic keys and has sufficient time at their disposal to leak the key even through such a slow communication channel.
As described in the small message theorem~\cite{moskowitz_covert_1994} even a few bits of data stolen over a long period can be enough to be harmful for a system if it is dependent on keeping them private.
The covert channel, like most other covert channels, has prerequisites regarding the underlying system and is therefore only usable in certain scenarios.
In our case, the covert channel fails if the load on the system by other processes is too high, if Turbo Boost is not enabled, or if the operating system does not place idle cores in deep sleep states to allow frequencies higher than the all-core turbo frequency.

The first, system utilization, is where our covert channel is significantly different from other frequency-based covert channels.
We demonstrate that our covert channel is still able to achieve usable throughput at up to 37.5\% of background load. 
As previously described, the average CPU utilization of modern data center is low enough that out covert channel is often still able to transmit data~\cite{garraghan_analysis_2013}. 
37.5\% background load is the maximum possible for the covert channel on our system, as in this scenario there are four cores active including that of the receiver.
As the lowest all-core turbo frequency is chosen at five active cores and above, any additional core will cause the final frequency drop.

Our covert channel is able to cope with higher background noise than other comparable frequency-based covert channels.
For example, Miedl et al.~\cite{miedl_frequency_2018} assume a completely idle system with only short bursts of background load.
Similarly, the prototype described by Khatamifard et al.~\cite{khatamifard_powert_2019} fails on Intel systems even if only one additional core is held active by other processes.

The other requirements, that Turbo Boost is enabled and that idle cores are placed in deep sleep states by the operating system, are likely fulfilled on most systems.
Disabling deep sleep states restricts the system to the all-core boost frequency and has significant impact on power consumption, while disabling Turbo Boost restricts the system to its base frequency, causing an even higher performance impact as described above.

\section{Related Work}
\label{sec:relwork}

While our covert channel exploits the power management of current Intel CPUs to provide a particularly robust covert channel in the face of background load caused by other processes, it is by far not the first covert channel based on CPU load.
In this section, we review existing load-based covert channels and describe similarities and differences.

\subsection{Load-Based Covert Channels}

The covert channel described in this paper is an example for a class of covert channels which transmit information by having the receiver observe the CPU load generated by the transmitter.
In contrast to our covert channel, most of these covert channels are easily mitigated without a negative impact on performance.

A basic example for such a covert channel is the original CPU load monitoring covert channel by Cioranesco et al.~\cite{cioranesco_communicating_2013} where the receiver directly observes CPU load via the sysstat \texttt{sar} command.
Access control by the operating system can be used to deny access to such commands which trivially mitigates the covert channel.
Another example is the thermal covert channel described by Masti et al.~\cite{masti_thermal_nodate}.
This channel uses the effect that CPU load increases the temperature not only of the loaded core but also of neighbor cores even if they are otherwise isolated by software.
If processes have user-space access to a temperature sensor of their current core, the heat propagation can be used to construct a inter-core covert channel and even a limited side channel.
However, as with other CPU load monitoring mechanisms provided by the OS, the operating system can deny user-space access to the temperature sensor.

\subsection{Frequency-Based Covert Channels}

Another metric influenced by the load on the system is the CPU frequency.
On modern systems, the frequency is affected by CPU load in two ways:
First, the operating system -- or, in some cases, the CPU itself -- lowers the frequency when load is low to save energy.
Second, the CPU increases the frequency above its nominal frequency when power headroom is available, as for example implemented by Intel Turbo Boost.

A covert channel exploiting the former has been described by Alagappan et al.~\cite{alagappan_dfs_2017}.
Their covert channel either manually lets the application set the frequency if the operating system allows -- which can be trivially denied through access control -- or varies CPU load to which the Linux frequency governor reacts by reducing the CPU frequency when possible.
At the receiver's side, the CPU frequency is either determined through the sysfs file system -- which can be trivially denied -- or through a timing loop similar to our covert channel.
On recent systems, however, the latter does not provide a cross-core covert channel anymore.
As individual cores can have different frequencies when operating below the system's turbo frequency, measurement on one core does not reflect another core's frequency.

Other approaches use the same mechanisms.
For example, the approach described by Miedl et al.~\cite{miedl_frequency_2018} uses varying CPU load to send information and uses a timing loop to directly measure CPU frequency, whereas the approach described by Benhani and Bossuet~\cite{benhani2019dvfs} directly manipulates the CPU clock to achieve significantly higher bit rates.
These approaches therefore suffer from the same limitations.
As our covert channel does not rely on specific operating system frequency management and provides cross-core communication in a lot of cases, the approach complements the existing approaches by providing a covert channel in situations where neither of them are viable.

The effect exploited by our covert channel is that modern CPUs implement boost frequencies to provide increase performance when possible.
As modern systems try to utilize available power headroom to maximize performance, increasing load on one core means that the system cannot sustain its previous boost frequency and that other cores might have to reduce their frequency.
Khatamifard et al.~\cite{khatamifard_powert_2019} show how this effect can be used to construct a range of covert channels that cannot be prevented by the operating system without significant performance impact.
They also provide a formal model for the achievable throughput and describe a prototype for Intel and ARM systems.
On the Intel system, their prototype achieves a peak channel capacity of \SI{121.6}{\bps}, which is similar to our covert channel if we remove error detection and correction from our prototype.  
As described in the introduction, their covert channel is not particularly optimized for a certain system architecture, though.
Their prototype tries to maximize power consumption when activating transmitter cores, which is not necessary on systems such as the one used in our evaluation and which might make detection of the attack easier.
Also, their prototype fails to transmit information on a sufficiently cooled system if even just a single additional core is held active.
While assuming an idle system is common in literature, this situation is highly unlikely in many practical environments, with modern cloud environments achieving up to 56\% CPU utilization on average~\cite{garraghan_analysis_2013}.
We describe an optimized covert channel for Intel Turbo Boost and show how the adaptation to the system allows for a significantly higher tolerance of the CPU load caused by other processes.

\subsection{Other Covert Channels}

Besides the load-based covert channels described above, a plethora of other mechanisms have been used to transmit information.
For example, the CPU caches~\cite{percival2005cache,wu2014whispers}, the DRAM row buffers~\cite{pessl_drama_2016}, and CPU functional units~\cite{wang_covert_2006} are shared resources that can enable covert channels with significantly higher throughput than the covert channel described in this work.
Other covert channels often also can be prevented by other countermeasures, though -- in these cases, cache partitioning, memory channel isolation, and disabling of hyper-threading can be used to prevent resource sharing.
Our covert channel or the channels described above do not supersede, but rather complement other covert channels.
Depending on the targeted system and the countermeasures deployed on the system, different covert channels can be used.

\subsection{Error Correction}

In Section~\ref{sec:impl:error_correction}, we described error correction methods to increase the usable (i.e., error-free) throughput of our covert channel.
The choice of error correction mechanisms depends on the characteristics of the channel.
For cache-based covert channels, Maurice et al.~\cite{maurice_hello_2017} provide a detailed characterization.
They describe the types of errors and show that with the right error correction mechanisms reliable communication with high throughput is possible.
On the physical layer a Berger code is used as an error detection code, combined with retransmissions of broken packets and a Reed-Solomon code for error correction on the data link layer.

Although the underlying mechanisms of the channel are different, our approach has to cope with similar types of errors, except that in our case the parallel execution of other programs on other cores can cause noise to affect many bits in series.
Our approach does not use error-correcting codes as estimates showed that Reed-Solomon codes would reduce performance.
Instead, our approach only relies on CRC-16 checksums for data integrity and on a retransmission scheme for robust transmission.

\subsection{Countermeasures}

In Section~\ref{sec:counter}, we describe countermeasures against our covert channel.
While the countermeasures are in most cases specific to our approach in particular or at least to frequency-based approaches in general, there are parallels to countermeasures against other types of covert channels.

For example, Brumley~\cite{brumley2011covert} and Schmidt et al.~\cite{schmidt2015case} describe pollution of cache lines as a countermeasure against cache-based covert channels. 
We show that CPU load noise has a significant impact on the throughput of our covert channel and is therefore usable as a countermeasure.
In our case, artificial noisy CPU load is effective because it randomly reduces the frequency of the system.
However, for that reason it potentially has a significant performance impact. 

In the past, reducing the resolution of the available time sources has been suggested as a countermeasure against timing-based covert channels as well as side channels~\cite{martin2012timewarp}.
However, even in highly restricted environments such as web browsers, where the timer resolution is artificially lowered, it is possible to achieve high temporal accuracy either by exploiting other unorthodox timing sources or by recovering precise timing information from the intentionally imprecise timer~\cite{schwarz2017fantastic}.

Countermeasures against cache-based covert channels include static partitioning of shared caches~\cite{wang2007new} or the use of randomized prefetchers in order to disguise memory accesses\cite{fuchs2015disruptive}.
The frequency covert channel equivalent of these techniques would be a power management mechanism that limits selection of turbo frequencies to subsets of the total cores and randomizes frequency selection.
No current hardware, however, implements such mechanisms.

\section{Conclusion}
\label{sec:conclusion}

Covert channels allow attackers to transfer information even if the access policy of the system prohibits the flow of information.
Previous work has identified changes to the CPU frequency as a method to transfer information between different programs either on the same core or on different cores.
However, existing approaches are either easily thwarted by the operating system at little to no cost or fail when the system experiences significant CPU load.

In this paper, we describe a covert channel specifically optimized to exploit Intel Turbo Boost.
Our transmitter varies the number of active cores, and our receiver detects the resulting changes to the maximum boost frequency.
By using multiple processes in the transmitter, we are able to increase the load swing so that the resulting covert channel can tolerate significant CPU load from other processes.
On an idle system, our prototype achieves a throughput of \SI{61}{\bps} between two processes on an idle system and \SI{43}{\bps} with 25\% utilization.
Even on a system with 37.5\% background load, our prototype is still able to successfully transmit \SI{12}{\bps} of data.
We show how available countermeasures against the covert channel have a significant impact on performance.

\bibliographystyle{plain}
\bibliography{bibliography.bib}

\end{document}